\documentclass[osajnl,twocolumn,showpacs,floatfix,bibnotes]{revtex4}

\usepackage{here}

\usepackage{graphicx}

\newcommand\pictc[5]{\begin{figure}
                       \centerline{
\includegraphics*[width=#1\columnwidth,height=0.7\textheight,keepaspectratio]{#3}}
                   \protect\caption{\protect\label{fig:#4} #5}
                    \end{figure}            }
\newcommand\pict[4][0.96]{\pictc{#1}{!tb}{#2}{#3}{#4}}
\newcommand\rpict[1]{\ref{fig:#1}}

\newcounter{Fig}

\begin{document}
\begin{sloppy}

\title{Observation of discrete gap solitons in binary waveguide arrays}

\author{Roberto Morandotti}
\affiliation{Universite  du Quebec, Institute National de la
Recherche Scientifique, Varennes, Quebec, Canada J3X 1S2}

\author{Daniel Mandelik}
\author{Yaron Silberberg}
\affiliation{Department of Physics of Complex Systems, Weizmann
Institute of Science, Rehovot 76100, Israel}

\author{Stewart Aitchison}
\affiliation{
Department of Electrical and Computer Engineering, University of Toronto, 
Canada M5S 3G4}

\author{Marc Sorel}
\affiliation{Department of Electrical and Electronic Engineering, University of Glasgow, Glasgow G12 8QQ, Scotland}

\author{Demetrios N. Christodoulides}
\affiliation{School of Optics/CREOL, University of Central Florida, Florida 32816-2700, USA}

\author{Andrey A. Sukhorukov}
\author{Yuri S. Kivshar}
\affiliation{Nonlinear Physics Group and Centre for Ultra-high
bandwidth Devices for Optical Systems (CUDOS), 
RSPhysSE, Australian National University,
Canberra, ACT 0200, Australia}

\begin{abstract}
We report on the first experimental observation of discrete gap
solitons in binary arrays of optical waveguides. We observe the
soliton generation when the inclination angle of an input beam is
slightly above the Bragg angle, and show that the propagation
direction of the emerging gap soliton depends on the input power
as a result of an inter-band momentum exchange.
\end{abstract}


\maketitle

Nonlinear periodic photonic structures such as arrays of optical waveguides
attracted a lot of interest due to the unique ways they offer for controlling light. 
Periodic modulation of the refractive index breaks the translational invariance and produces effective discreteness in a continuous system, 
opening up many novel possibilities for manipulating light propagation, including light localization in the form of {\em discrete optical solitons}~\cite{Christodoulides:1988-794:OL, Eisenberg:1998-3383:PRL}.

Recently, it was suggested that a novel type of discrete optical
solitons, the so-called discrete gap solitons, can be generated in
the binary waveguide arrays, by a single inclined
beam~\cite{Sukhorukov:2002-2112:OL} or two input
beams~\cite{Sukhorukov:2003-2345:OL}. The binary arrays of optical
waveguides are specially engineered photonic structures consisting
of periodically alternating wide and narrow waveguides [see an
example in Fig.~\rpict{bloch}(a)]. Discrete gap solitons in such
structures can be considered as a nontrivial generalization of the
spatial gap solitons recently observed in the waveguide
arrays~\cite{Mandelik:2004-93904:PRL} and optically-induced
photonic lattices~\cite{Neshev:nlin.PS/0311059:ARXIV}. Gap
solitons in binary arrays are associated with the fundamental
modes strongly confined in narrow waveguides, which lowers the
power threshold for the soliton excitation, and may result in
reduced radiation losses compared to the gap solitons based on
radiation or higher-order bound modes in the conventional arrays.
In this Letter, we report on the first experimental observation of
discrete gap solitons in binary arrays of optical waveguides and
demonstrate {\em a novel method of efficient steering} of gap
solitons based on the inter-band momentum exchange.

We investigate spatial beam self-action and soliton formation in
the fabricated  etched arrays of the $5$~mm long AlGaAs waveguides
with the effective refractive index contrast $0.0035$ (see
Ref.~\onlinecite{Eisenberg:1998-3383:PRL}). The binary arrays are
made of wide (4~$\mu m$) and narrow (2.5~$\mu m$) waveguides with
4~$\mu m$ edge-to-edge spacing, and accordingly the full period is $d=$14.5$\mu m$ as illustrated in Fig.~\rpict{bloch}(a). The beam propagation in this structure can
be described by the normalized nonlinear Schr\"odinger equation,
$   i \frac{\partial E}{\partial z}
   + D \frac{\partial^2 E}{\partial x^2}
   + \nu(x) E + |E|^2 E = 0$,
where $E(x,z)$ is the normalized envelope of the electric field,
$x$ and $z$ are the transverse and propagation coordinates
normalized to the characteristic values $x_s=1 \mu m$ and $z_s=1
mm$, respectively, $D = z_s \lambda / (4 \pi n_0 x_s^2)$ is the
beam diffraction coefficient, $n_0=3.3947$ is the average medium
refractive index, $\lambda=1.5\,\mu m$ is the vacuum wavelength,
$\nu(x) = 2 \Delta n(x) \pi n_0 / \lambda$, and $\Delta n(x)$ is
the effective modulation of the optical refractive index. 

\pict{fig01.eps}{bloch}{ (a)~Normalized refractive index  in a binary waveguide array (gray) and Bloch-wave profiles at the top of first (solid) and second (dashed) bands. (b,c)~Dispersion of Bloch waves and the Bloch-wave excitation coefficients vs. the transverse wave-number, for the
first (solid) and second (dashed) bands. Fourier spectrum of an
input beams with 20\% larger then Bragg incident angle is shown in
(c) with shading. (d)~Characteristic profile of a discrete gap
soliton in the binary array. }

In the waveguide arrays, the wave spectrum possesses a band-gap
structure with gaps separating bands of the continuous spectrum
associated with the spatially extended Bloch waves of the form,
$E(x,z) = \psi(x) \exp(i k x + i \beta z)$, where the propagation
constant $\beta$ is proportional to the wave-vector component
along the waveguides. In Fig.~\rpict{bloch}(b), we plot the
Bloch-wave dispersion curves and mark {\em two types of spectrum
gaps} appearing due to (i)~total internal reflection and
(ii)~Bragg-type scattering.

\pict{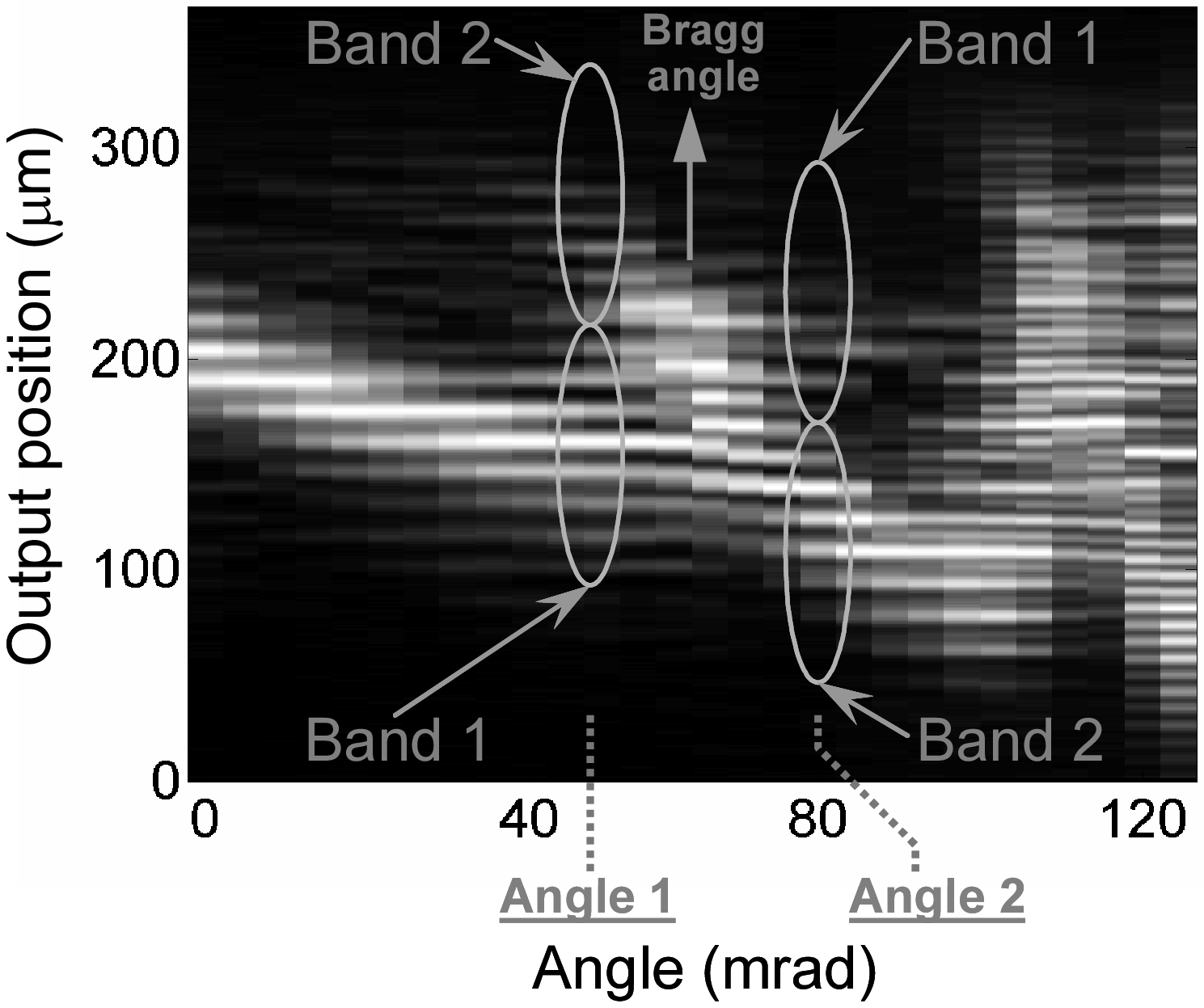}{scatterExper}{Experimentally measured output
intensity distribution vs. the inclination angle of the input Gaussian
beam. }

\pict{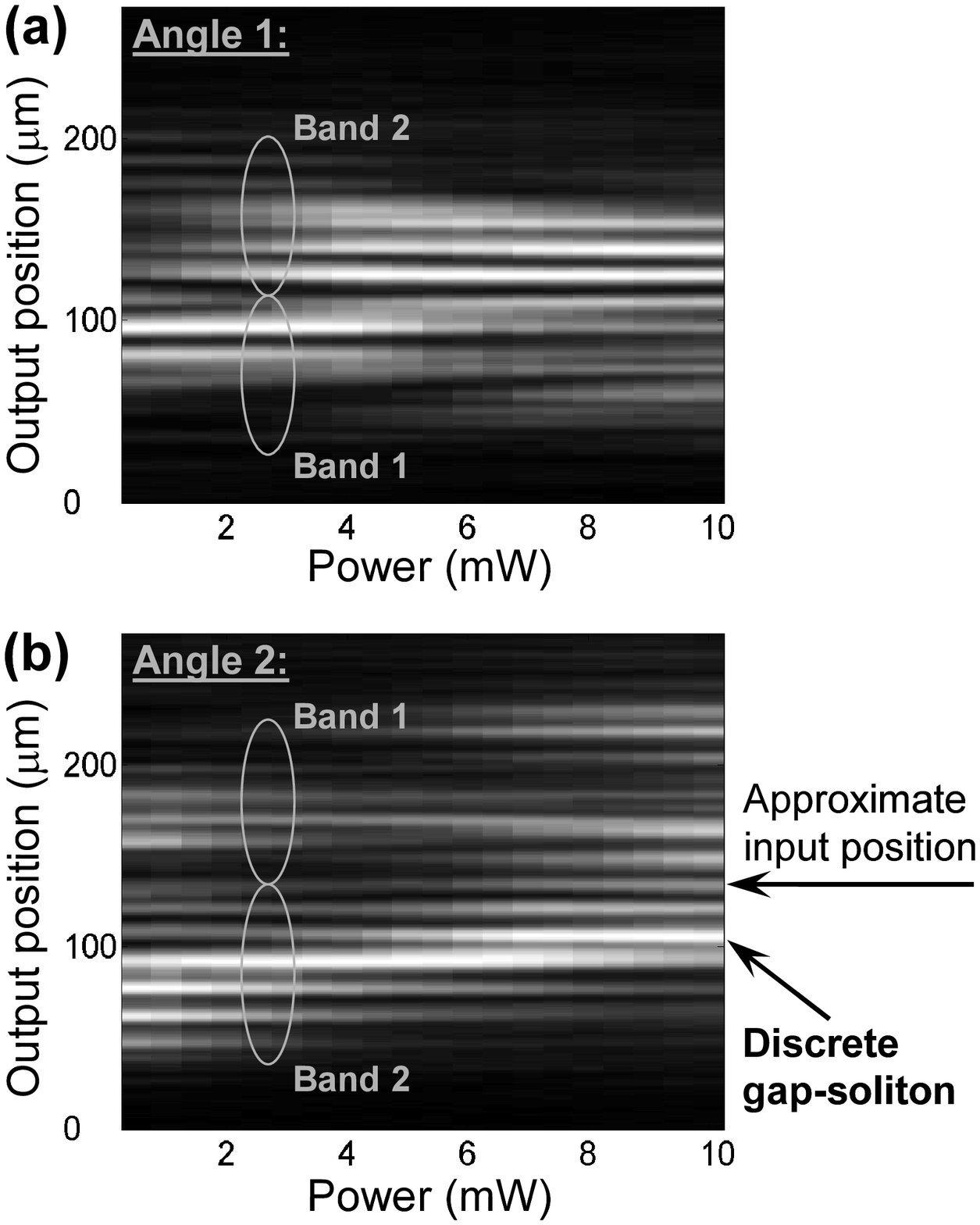}{nonlExper}{ (a,b)~Experimentally measured intensity
distribution at the output vs. the average input power for 
incident angles marked 1 and 2 in Fig.~\rpict{scatterExper},
respectively. }

When the medium nonlinearity is self-focusing, the discrete gap
solitons may appear near the upper edge of the second
band~\cite{Sukhorukov:2003-2345:OL}. One of the distinguishing
features of such solitons is their localization at narrow
waveguides, as shown in Fig.~\rpict{bloch}(d). As was suggested in
Ref.~\onlinecite{Sukhorukov:2002-2112:OL}, the gap solitons can be
excited by an input Gaussian beam which is inclined slightly above
the Bragg angle. Each Fourier component of an incident beam
excites a superposition of Bloch waves belonging to different
bands, according to the values of excitation coefficients $C_n$
plotted in Fig.~\rpict{bloch} (see details in Ref.~\onlinecite{Sukhorukov:2004-93901:PRL}). If the input beam spans several periods of the underlying lattice (we
choose the 45~$\mu m$ FWHM beam) and its spectrum is narrow, then
the excitation of the second band becomes dominant [see
Fig.~\rpict{bloch}(c)], facilitating the formation of a gap
soliton. Although fully controlled generation of gap solitons can
be achieved only in a two-beam excitation
scheme~\cite{Feng:1993-1302:OL, Sukhorukov:2003-2345:OL,
Mandelik:2004-93904:PRL, Neshev:nlin.PS/0311059:ARXIV}, the
single-beam approach is simpler to implement. Additionally, the
simultaneously excited Bloch waves of the first band move in the
opposite direction, resulting in
``self-clearing'' of the gap soliton, and allowing to control the
soliton velocity by changing the input beam
intensity~\cite{Sukhorukov:2002-2112:OL}. We note that all
Bloch-wave components tend to trap together into a multi-band
breather if a normally incident narrow beam is
used~\cite{Mandelik:2003-253902:PRL}, and complex Bloch-wave
interaction occurs for the inclined narrow
beams~\cite{Sukhorukov:2004-93901:PRL}.

\pict{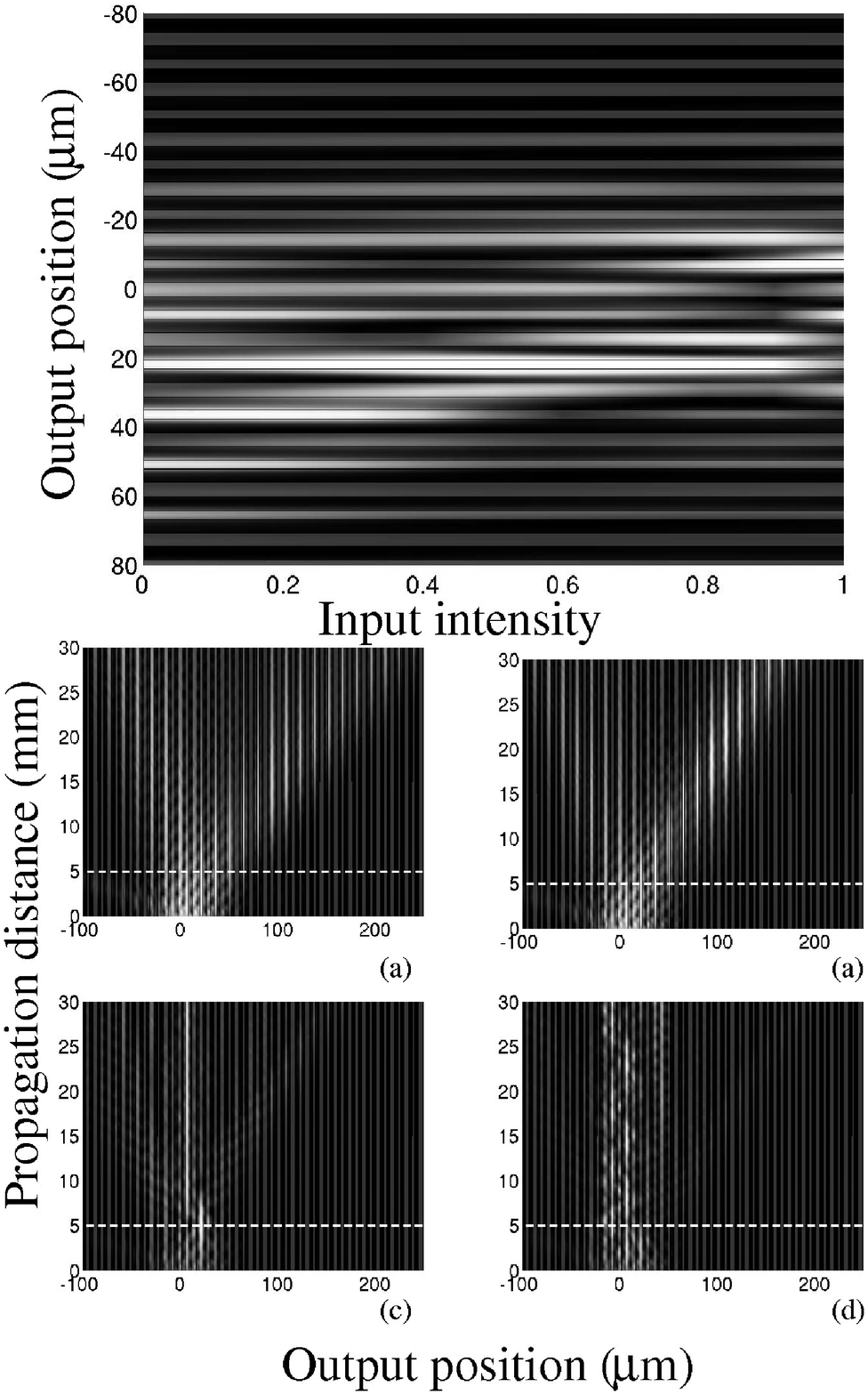}{nonlTheor}{ Top: Numerically calculated
output field intensity vs. the peak input intensity ($I_0$) for the
beam inclination 20\% above the Bragg angle, corresponding to
Fig.~\rpict{nonlExper}(b). (a-d): Beam propagation along the
binary waveguide array for different input intensities:
(a)~diffraction in linear regime ($I_0=0$); (b)~formation of a
moving discrete gap soliton at $I_0=0.1$; (c)~excitation of slow
or stationary gap solitons at $I_0=0.5$; (d)~soliton
destabilization above a critical power threshold ($I_0=1$). Dashed
line marks the boundary of experimental sample. }

In our experiments, first we study the linear beam propagation (see Ref.~\onlinecite{Eisenberg:1998-3383:PRL} for description of experimental set-up).
The dependence of the output field distribution on the input angle
is shown in Fig.~\rpict{scatterExper}. We note that, in a binary
waveguide array, the Bloch waves corresponding to the first and
second bands can be easily distinguished, since they are primarily
localized at wide and narrow waveguides, respectively. We find
that the band-2 excitation becomes dominant above the Bragg angle,
in full agreement with the theoretical predictions (higher bands become excited as the angle is further increased).

Next, we analyze nonlinear self-action of beams with the
inclination angles below and above the Bragg angle, as marked in
Fig.~\rpict{scatterExper}. In the former case, the Bloch waves
associated with the bottom of the first band and experiencing {\em
anomalous diffraction} are primarily excited. This results in beam
self-defocusing as nonlinearity
grows~\cite{Morandotti:2001-3296:PRL}. Indeed, we register
broadening of the output beam with the increase of the laser
power, see Fig.~\rpict{nonlExper}(a). 

Strong beam self-focusing
and gap-soliton formation are observed when the input inclination
angle is above the Bragg angle [see Fig.~\rpict{nonlExper}(b)],
and the gap solitons are localized at narrow waveguides.
Additionally, the output soliton position depends on the input
power.
Numerical results for the beam propagation 
[Fig.~\rpict{nonlTheor}, top] agree well with the experimental
data. In Figs.~\rpict{nonlTheor}(a-d) we show the beam dynamics inside the waveguide array structure for various input intensities.
These simulations demonstrate that  the soliton velocity and,
accordingly, the soliton output position are defined largely at
the initial stage, where the counter-propagating Bloch waves of
the first and second bands interact strongly with each other. This
inter-band momentum exchange is predominantly responsible for the
soliton steering, which may find applications for a simple
realization of all-optical switching.

In conclusion, we have generated 
discrete gap solitons in binary waveguide arrays and presented an experimental evidence of the inter-band momentum exchange observed in the form of a power-dependent soliton steering. %Our results are in a perfect agreement with earlier theoretical predictions~\cite{Sukhorukov:2002-2112:OL}, and they open a road for deeper experimental studies of other intriguing properties of gap solitons in engineered periodic structures.

\end{sloppy}
\end{document}